\documentclass{article}

\usepackage{graphicx}
\usepackage{latexsym}
\usepackage{amsfonts}
\usepackage{url,hyperref}
\usepackage{bm}

\def\abstract#1{\vskip 7mm 
        \begin{center}{\large Abstract}\par \smallskip
                \begin{minipage}[c]{12cm}
                        \small #1
                \end{minipage}
        \end{center}
}
\def\title#1{\begin{center}{\Large\bf #1}\end{center}}
\def\author#1{\vskip 5mm \begin{center}{#1}\end{center}}
\def\address#1{\begin{center}{\it #1}\end{center}}
\makeatletter
\@ifundefined{lesssim}{}{}
\@ifundefined{gtrsim}{}{}
\def\vereq#1#2{\lower3pt\vbox{\baselineskip1.5pt \lineskip1.5pt
\ialign{$\m@th#1\hfill##\hfil$\crcr#2\crcr\sim\crcr}}}
\makeatother

\begin{document}

\newcommand{\vp}{\varphi}
\newcommand{\nn}{\nonumber\\}
\newcommand{\beq}{\begin{equation}}
\newcommand{\eeq}{\end{equation}}
\newcommand{\bed}{\begin{displaymath}}
\newcommand{\eed}{\end{displaymath}}
\def\bea{\begin{eqnarray}}
\def\eea{\end{eqnarray}}
\newcommand{\veps}{\varepsilon}

\title{%
  Fermion evaporation of a black hole off a tense brane
  \smallskip \\
  {\large Proceedings for the JGRG17 international conference Nagoya, Japan, December 2007}
}
\author{%
H.~T.~Cho\footnote{htcho"at"mail.tku.edu.tw},
A.~S.~Cornell\footnote{cornell@ipnl.in2p3.fr},
Jason~Doukas\footnote{j.doukas"at"physics.unimelb.edu.au},
Wade~Naylor\footnote{naylor"at"se.ritsumei.ac.jp}
}
\address{%
$^1$ Department of Physics, Tamkang University, Tamsui, Taipei, Taiwan, Republic of China
\\
$^2$ Universit\'e Lyon 1, Institut de Physique Nucl\'eaire de Lyon, France
\\
$^3$ School of Physics, University of Melbourne, Parkville, Victoria 3010, Australia
\\
$^4$ Department of Physics, Ritsumeikan University, Kusatsu, Shiga 525-8577, Japan
}

\abstract{
Using the WKBJ approximation we obtain numerical plots of the power emission spectrum for the evaporation of massless bulk Dirac fermions from six dimensional black holes off a tense 3-brane with codimension two. We also present the multiplicity factors for eigenvalues of the deficit four sphere and show that these reduce to the usual case in the tenseless limit.}

\section{Introduction}

\par One of the most distinct predictions of large extra-dimensional models \cite{ADD} is the production of black holes (BHs) at particle accelerators such as the LHC \cite{BHacc}. In these models the standard model (SM) fields are restricted to motion along a 3-brane in the extra dimensions, while the gravitational field can propagate isotropically. By readjusting the fundamental cut-off scale one can simultaneously resolve the gauge-hierarchy problem and give some explanation for the observed weakness of gravity relative to the other forces. Recently, a metric describing a BH located on
a 3-brane with finite tension, embedded in locally flat
6-dimensional (6D) spacetime was discovered \cite{KK}. Before this, no exact solution that incorporated brane tension was known and the effect of brane tension on the observational signatures of mini BHs was largely ignored.

\par While early incarnations of large extra dimensional models assumed that only gravitational fields existed in the space off the brane (known as the \textit{bulk}), over time more elaborate scenarios were devised that required other bulk fields. Split fermion models \cite{Split} are examples of this kind. In split fermion models proton decay inducing operators can be suppressed while simultaneously giving the correct SM mass hierarchies. The result is achieved by using a kink configuration of a bulk scalar field to localise quarks and leptons and the left- and right- chiral components of the fermion fields to different locations in the higher dimension(s). In supersymmetric versions of this idea the scalars that localise the bulk fields will also have bulk fermionic superpartners.

\par Having a larger number of degrees of freedom propagating in the bulk would lower the probability of witnessing a BH event at the LHC i.e., evaporation into bulk modes would reduce the amount radiation seen from the brane thereby lowering the chance of identifying a BH event \footnote{While the missing energy may be some evidence for extra dimensions it would not conclusively identify a BH event.}. Regardless of which model you prefer, there is clearly an imperative to determine precisely how bulk modes effect the Hawking emission spectrum so that in the event that a mini BH is observed the correct extra dimensional model may be able to be inferred.

\par In what follows we study the effect of brane tension on the BH emission spectrum of massless bulk fermions. The analogous emission rates for scalar, gauge boson and graviton bulk fields were
calculated in \cite{Dai}, however as we will be using the WKBJ approximation our treatment differs somewhat to theirs.

\par The metric for a black hole residing on a tensional 3-brane embedded in a six-dimensional spacetime is \cite{KK,CWSSiopsis}:
\beq
\label{metric}
 \mathrm{d}s^{2}=-f(r)dt^{2}
+\frac{dr^{2}}{f(r)}+ r^2 d\Omega_4^2 \ \ , \ \ \ \
f(r)=1-\left(\frac{r_{H}}{r}\right)^{3}
 \eeq
 where the radius of the horizon is given by
\beq\label{eqn:tenseradius}
r_H=\left(\frac{\mu}{b}\right)^{1/3} \,\qquad\qquad\, \mu \equiv \frac{M_{BH}}{4\pi^2M_*^4}
\eeq
and $M_{BH}$ is the mass of the black hole. The parameter $b$ is a measure of the conical
deviation from a perfect sphere and has the following angle element:
\beq
 d\Omega_4^2 = d\theta^{2}_3+\sin^{2}\theta_3
\left(d\theta_2^2+\sin^{2}\theta_2 \left(d\theta_1 ^{2}+
b^{2}\sin^{2}\theta_1 d\phi ^{2}\right) \right), \ \ \ \ 0 < b \le 1 .
\label{anglel}
\eeq
For $b=1$ this is the line element of the unit sphere $S^4$ and corresponds to zero brane tension. In the case of non-vanishing brane tension the parameter $b<1$ is a measure of the deficit angle about an axis parallel to the 3-brane in the angular direction $\phi$, such that the canonically normalized angle $\phi'=\phi/b$ runs over the interval $[0,2\pi/b]$. It can be expressed in term of the brane tension $\lambda$ as:
\beq
\label{eqb} b = 1 - \frac{\lambda}{4\pi M_*^4} \,,
\eeq
where $M_*$ is the fundamental Planck constant of six-dimensional gravity. As can be seen the tension of the brane ($b\to 0$) increases the radius of the horizon.

\par The Dirac operator on this metric is solved by use of the conformal transformation:
\begin{eqnarray}
g_{\mu\nu} & \rightarrow & \overline{g}_{\mu\nu}=\Omega^{2}g_{\mu\nu} , \\
\psi & \rightarrow & \overline{\psi}=\Omega^{-5/2}\psi , \\
\gamma^{\mu}\nabla_{\mu}\psi & \rightarrow & \Omega^{7/2} \overline{\gamma}^{\mu}\overline{\nabla}_{\mu}\overline{\psi} ,
\end{eqnarray}
where $\Omega=1/r$, the metric in equation (\ref{metric}) then separates into a $t-r$ part and a deficit 4-sphere part:
\beq
d\overline{s}^{2} = \frac{1}{r^2}\left(-f(r)dt^{2} + \frac{1}{f(r)}dr^{2}\right) + d\Omega^{2}_{4}.
\eeq
The massless Dirac equation, $\overline{\gamma}^{\mu}\overline{\nabla}_{\mu}\overline{\psi}=0$,  can then be written:
\begin{eqnarray}
\left[ \left( \overline{\gamma}^{t}\overline{\nabla}_{t} + \overline{\gamma}^{r}\overline{\nabla}_{r} \right) \otimes 1 \right] \overline{\psi} + \left[ \overline{\gamma}^{5} \otimes \left( \overline{\gamma}^{a} \overline{\nabla}_{a}\right)_{S_{4}} \right] \overline{\psi} = 0 , &
\end{eqnarray}
where $(\overline{\gamma}^{5})^{2}=1$ and $\overline{\psi}=r^{5/2}\psi$. Note that from this point on we shall change our notation by omitting the bars.
\par The eigenvalues, $\kappa$, for the eigenspinors of the deficit $4$-sphere,
\begin{equation}\label{eqn:angularEigen}
\left( \gamma^{a}\nabla_{a} \right)_{S_{4}}\chi_{l}^{(\pm)} = \pm i \kappa \chi_{l}^{(\pm)},
\end{equation}
where found in \cite{CCDN} and are given by
\beq
\kappa(n,m)=n+2+|m|\left({1\over b}-1\right),
\eeq
where  $n=0,1,2,\dots$ and $m=\pm 1/2,\pm 3/2, \dots, \pm (n+1/2)$. After a little algebra \cite{CCDN}, the radial part of the Dirac equation reduces to a Schr\"odinger-like equation in the tortoise coordinate $r_{*}$:
\beq
\left(-\frac{d^2}{dr_{*}^{2}}+V_{1}\right)G=E^{2}G \,\,\, ,
\label{gequation}
\eeq
where $dr=f(r) dr_{*}$, and the potential is given by $V_1(r) = \kappa^2 {f\over r^2}+\kappa
f\frac{d}{dr}\left[\frac{\sqrt{f}}{r}\right]$. 

\section{Absorption probability and emission rates}
\begin{figure}[h]
\centering
\includegraphics[scale=.6]{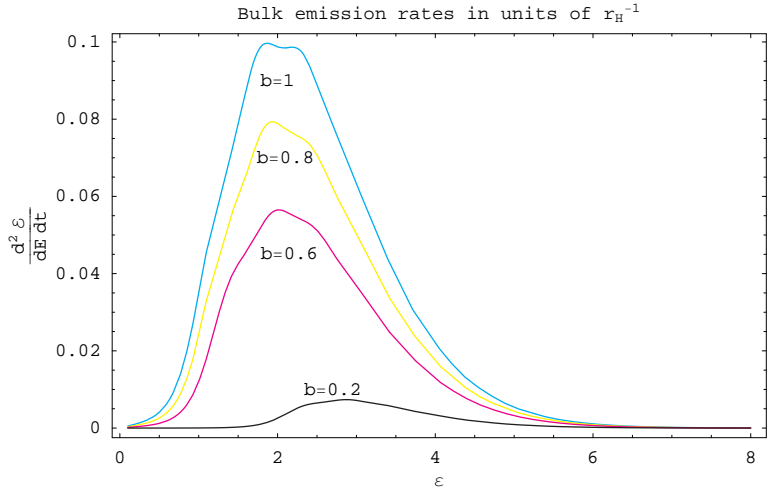}
\caption{We display the results of third order WKBJ, which show the variation of the Hawking radiation spectrum with brane tension. Increasing brane tension (decreasing $b$), while holding $r_H$ and $M_*$ fixed, results in a reduction of the emitted power.}
\label{fig:tenseemission}
\end{figure}
\par When the BH perturbation equation takes the Schr\"odinger form as in equation (\ref{gequation}) an adapted form of the WKBJ method \cite{WG} can be employed to find the absorption probability. The absorption probability is written:
\beq
|{\cal A}_\kappa(E)|^2 = {1\over 1+e^{2S(E)}} \,\,\, ,
\label{IMabprob}
\eeq
where $S(E)$ is calculated to third order in \cite{CCDN,WG}.
In our case it will be convenient to make a change of variables to $x=Er$ \cite{Cornell}. This leads to the following form of the potential:
\beq Q(x_*)= 1 - \kappa^2 {f\over x^2} - \kappa f \frac{d}{dx}
\left[ \frac{\sqrt{f}}{x}\right] \,, \qquad \quad f(x) = 1 -
\left( \veps\over x\right)^{d-3},\label{definitionofq} \eeq where $E^2 Q(x_*) = E^2 - V_1 $, and $\veps = r_H
E$. As such, the Schr\"{o}dinger
equation (\ref{gequation}) becomes: \beq
\left(\frac{d^2}{dx_{*}^{2}}+Q\right)G=0 \,\,\, . \eeq
The emission rate for a massless fermion from a BH is related to the cross-section by a $d^{5}k$ dimensional momentum integral times a fermionic thermal temperature distribution:
\beq
{d {\cal E}\over dt} = \sum_{\lambda,E} \sigma_{\lambda,E} {E \over e^{E \over T_H}+1} {d^{5}k \over (2\pi)^{5}} \,\,\, , \label{EmX}
\eeq
where $T_H$ is the Hawking temperature, $\sigma_{\lambda,E}$ are the greybody factors and the sum is a generic sum over all angular momentum and momentum variables. The greybody factor can be related to the absorption probability by considering the results of reference \cite{Cardoso}:
\beq
\sigma_{\lambda,E} = {1\over 2 \Omega_{4}} \left( 2\pi\over E\right)^{4} \sum_\kappa D_\kappa
|{\cal A}_\kappa(E)|^2 \,\,\, . \label{Xsec}
\eeq
The degeneracy, $D_\kappa$, of eigenspinors for the deficit four sphere is found to be:
\bea
D_{\kappa}(n,|m|)&=&\sum_{n_2=|m|-1/2}^n2\sum_{n_1=|m|-1/2}^{n_2}2,\nonumber\\
&=&2\left(n-|m|+\frac{5}{2}\right)\left(n-|m|+\frac{3}{2}\right).
\eea
Summing over $|m|$ reproduces to the degeneracy calculated in \cite{Camporesi} for the regular four sphere:
\bea
D_4(n)&=&\sum_{|m|=1/2,\dots,n+1/2}2\left(n-|m|+\frac{5}{2}\right)\left(n-|m|+\frac{3}{2}\right),\nonumber\\
&=&\frac{2}{3}(n+1)(n+2)(n+3),
\eea
Given that 
$\int d^{5}k=\int \Omega_{4} E^{4} dE$, 
and using the fact that the Hawking temperature is $T_H=5/(4\pi r_H)$, we obtain:
\beq {d^2{\cal E}\over dE dt} = {1 \over \pi r_H}
\sum_{\kappa>0}{\veps \over e^{4\pi \veps \over 5} + 1}
D_{\kappa} |{\cal A}_{\kappa}(\veps)|^2 \,\,\, . \label{Miss} \eeq

\section{Results and Conclusion}
\par The power emission for various values of the tension parameter $b$ are overlaid in figure \ref{fig:tenseemission}. In plotting the graphs as a function of epsilon we have implicitly assumed that the horizon radius is fixed, i.e., $\epsilon=Er_H$. Since $r_H=(\frac{\mu}{b})^{1/3}$ depends on $b$, the ratio $\frac{\mu}{b}$ is also fixed, this can be done, see equation (\ref{eqn:tenseradius}), by fixing the fundamental scale, $M_*$, and changing the mass of the black hole proportionally with $b$.
\par We find that for increased tension ($b\rightarrow 0$) the emitted power due to Hawking radiation is reduced. These results are consistent with the those made for the integer representations of the Poincare group in \cite{Dai}, and completes a missing piece of the picture of emission from BHs off tense branes in being the first time that the power emission spectrum from massless spin 1/2 fields has been calculated.


\end{document}